\documentclass[lettersize,journal]{IEEEtran}
\usepackage{amsmath,amsfonts}
\usepackage{algorithmic}
\usepackage{algorithm}
\usepackage{array}
\usepackage[caption=false,font=normalsize,labelfont=sf,textfont=sf]{subfig}
\usepackage{textcomp}
\usepackage{stfloats}
\usepackage{url}
\usepackage{verbatim}
\usepackage{graphicx}
\usepackage{cite}
\usepackage{bm}
\usepackage{booktabs} 
\usepackage{amsmath} 

\usepackage[colorlinks=false]{hyperref}
\hypersetup{
    citebordercolor={0 1 0},    
    linkbordercolor={1 0 0},    
    pdfborder={0 0 1}           
}

\usepackage{booktabs}   
\usepackage{multirow}   
\usepackage[most]{tcolorbox}   
\usepackage[table]{xcolor} 
\usepackage{array}

\usepackage{pifont}

\newtcolorbox{keynotebox}{
    colback=yellow!10,
    colframe=orange!75!black,
    arc=4pt,
    boxrule=1pt,
    left=6pt, right=6pt,
    top=4pt, bottom=4pt,
    drop shadow,
    width=\linewidth
}

\begin{document}

\title{WaferTrans: Enabling IOMMU-free Distributed Virtual Address Translation for Wafer-scale GPUs}

\author{Xinru Tang, Jingxiang Hou, Guanghong Wu, Yang Hu, and Shouyi Yin, \IEEEmembership{Fellow, IEEE}

\thanks{Manuscript received September 4, 2026.}
\thanks{The authors are with the School of Integrated Circuits and Systems, Tsinghua University, Beijing 100084, China. E-mail: \{tangxr23, wugh25, houjx22\}@mails.tsinghua.edu.cn, \{hu\_yang ,yinsy\}@tsinghua.edu.cn}
}

\markboth{IEEE Journal of \LaTeX\ Class,~Vol.~12, No.~6, February~2024}%
{Shell \MakeLowercase{\textit{et al.}}: A Sample Article Using IEEEtran.cls for IEEE Journals}

\maketitle
\begin{abstract}
Wafer-scale GPUs (WSGs) provide sufficient on-wafer bandwidth to make near-lossless Unified Memory feasible. However, existing designs still rely on a CPU-IOMMU to translate remote virtual-address accesses. This centralized mechanism scales poorly to tens of GPU dies: translation requests must traverse costly off-wafer hierarchies and contend for limited CPU-side resources, making address translation a critical bottleneck. We propose WaferTrans, an IOMMU-free distributed virtual-address translation design for WSGs. WaferTrans introduces PTE Presence Consistency (PTE-PC), a lightweight consistency model that tracks PTE insertions and removals, and equips each GPU with a PTE Presence Directory (PPD) that locates the GPU holding a requested PTE. It further employs a distributed PTE-PC mapping and a cooperative query mechanism to localize PTE-PC maintenance while preserving complete lookup coverage. Together, these mechanisms enable the GPU array to resolve remote translations within the wafer, eliminating its dependence on the CPU-IOMMU. Compared with the SOTA Trans-FW design, WaferTrans improves performance by 2.5x on average.
\end{abstract}


\section{Introduction}

Modern GPU systems pursue near-linear scaling by aggregating more devices into a Unified Memory (UM) system, where remote data access should approach the cost of local access. Achieving this goal requires per-device interconnect bandwidth comparable to memory bandwidth, i.e., at least $\frac{N-1}{N} \approx 1$ of local memory bandwidth~\cite{arunkumar2017mcm}. Existing platforms still fall short. For example, NVIDIA B200 provides 8~TB/s memory bandwidth but only 0.9~TB/s interconnect bandwidth, yielding an interconnect-to-memory ratio of 0.225. Consequently, conventional multi-GPU systems cannot sustain ideal UM performance as they scale.

\begin{figure}[t]
    \centering
    \includegraphics[width=1.0\linewidth]{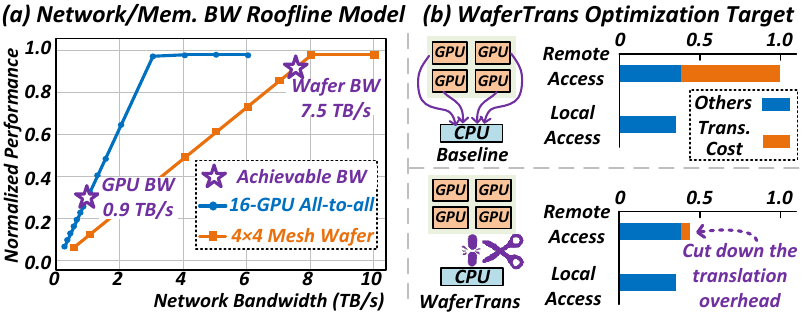}
    \caption{(a) Network/Memory bandwidth roofline model on idealized GEMM kernel--ignore translation overhead. (b) Optimization target of WaferTrans.}
    \label{fig:intro}
\end{figure}

Wafer-scale GPUs (WSGs), enabled by recent CoWoS advances, offer a more promising substrate~\cite{hu2024wafer,tang2026moentwine}. By integrating multiple GPU dies on a wafer-scale interposer, a WSG system provides a uniform, low-latency, high-bandwidth fabric across the GPU array. Reported on-wafer bandwidth can reach single-direction 7.5~TB/s~\cite{shih2025sow}, pushing the interconnect-to-memory ratio beyond 1 and making near-lossless UM feasible. Fig. \ref{fig:intro}(a) shows an idealized GEMM experiment without translation overhead, illustrating this potential: WSG performance reaches a saturation point because high on-wafer bandwidth removes remote data access as the bottleneck, yielding a clear advantage over conventional GPU systems.

However, current UM systems still rely on a CPU-IOMMU as a centralized address-translation engine. This dependency is ill-suited to WSGs: every remote translation request must traverse the wafer-to-CPU path and contend for limited CPU-side translation resources. The problem worsens as a wafer integrates tens of GPU dies. Trans-FW~\cite{li2023trans} adds a forwarding table (FT) at the IOMMU to direct requests to the GPU holding the target page table entry (PTE), but it cannot eliminate the costly off-wafer transfer; moreover, the centralized FT itself becomes congested under the large volume of requests generated by thousands of GPU SMs. Thus, address translation becomes the critical bottleneck, as shown in Fig. \ref{fig:intro}(b).

To address this problem, this work proposes \textbf{WaferTrans}, which decouples WSGs from CPU-centric translation and enables the GPU array to resolve translations cooperatively. The result is IOMMU-free virtual-address translation, unlocking full UM potential. Compared with the SOTA Trans-FW \cite{li2023trans} design, WaferTrans improves performance by 2.5x on average.

\section{Background \& Motivation}

\subsection{Introduction to Wafer-scale GPUs (WSGs)}

\begin{figure}[t]
    \centering
    \includegraphics[width=1.0\linewidth]{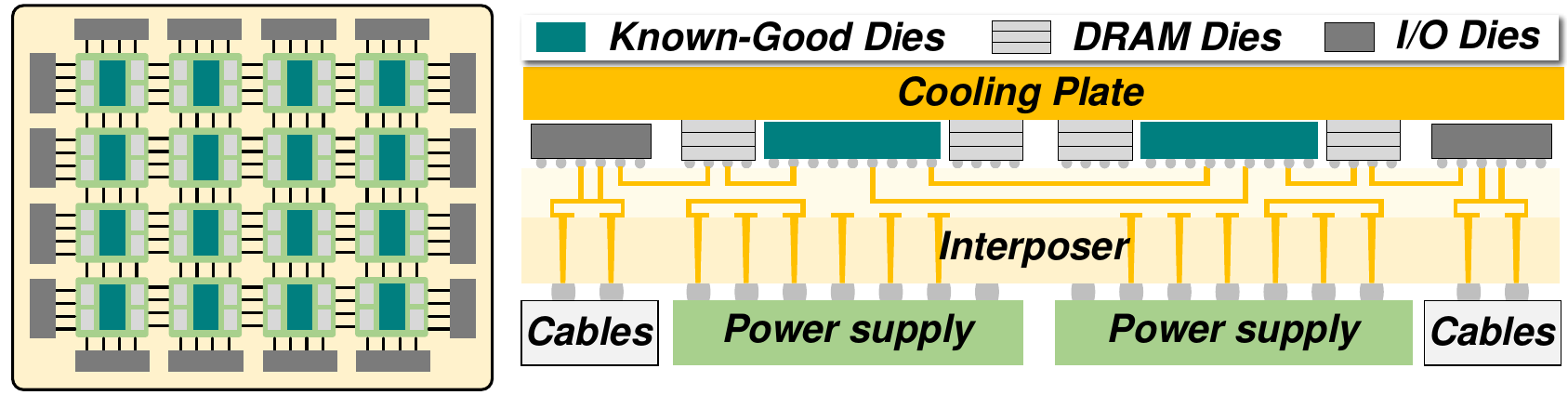}
    \caption{Top and cross-sectional views of a wafer-scale GPU (WSG) system.}
    \label{fig:wafer}
\end{figure}

Fig. \ref{fig:wafer} shows a typical WSGs\cite{hu2024wafer, tang2026moentwine,jiang2026wsccost}. Compute dies and DRAM modules are bonded onto a wafer-scale interposer, while I/O dies at the wafer periphery provide off-wafer connectivity. Because on-wafer links are short and traverse few packaging layers, they can deliver much higher bandwidth and lower latency than conventional off-package links. Meanwhile, signal-integrity limits make long, high-bandwidth links across many dies unattractive, so practical WSGs naturally favor a mesh-like topology. \textbf{\textit{We therefore view a WSG as a mesh-connected GPU cluster with high-performance interconnect.}}

\subsection{Address Translation on WSGs}

\begin{figure}[t]
    \centering
    \includegraphics[width=1.0\linewidth]{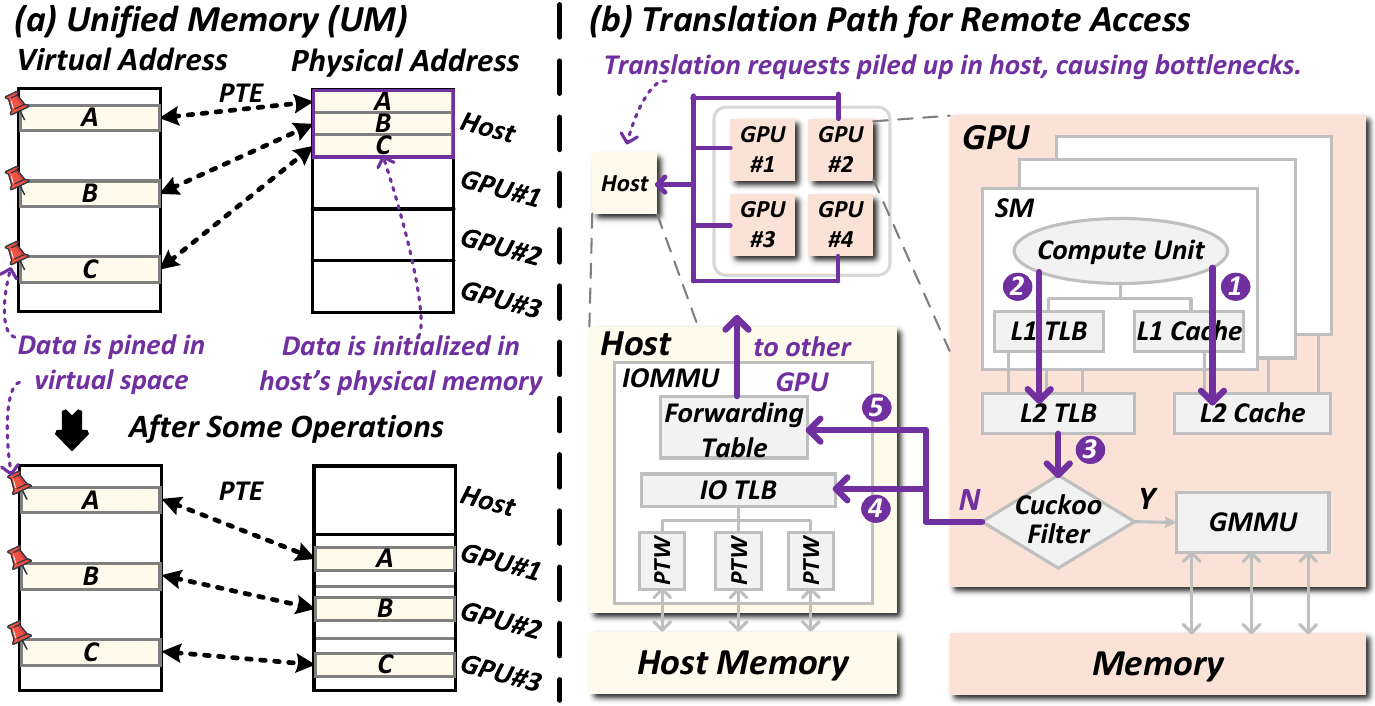}
    \caption{(a) Concept of a Unified Memory (UM) system. (b) Remote-access translation path in a CPU-centric architecture.}
    \label{fig:background}
\end{figure}

Fig.~\ref{fig:background}(a) summarizes Unified Memory (UM). UM exposes data through a single virtual address space, so a pointer is independent of the device currently storing the page. Physical placement changes over time: pages are initially backed by CPU memory and may migrate to GPU memory. Virtual-to-physical mappings are recorded in page table entries (PTEs). The host maintains the complete PTE set, whereas each GPU stores only PTEs for pages resident in its local memory. Before accessing a remote page, a GPU must obtain the corresponding PTE to determine its physical location; thus, remote accesses ultimately depend on CPU-side translation.

Fig.~\ref{fig:background}(b) illustrates the translation path for a remote access. {\large\ding{182}}An SM issues a memory request that misses in the local L1/L2 cache hierarchy and {\large\ding{183}}then in the local TLB hierarchy, which caches frequently used PTEs. {\large\ding{184}}A VPN-indexed cuckoo filter~\cite{li2023trans} checks whether the requested PTE is present locally. If the local lookup misses, {\large\ding{185}}the GPU forwards the translation request to the CPU, where the CPU-IOMMU performs a page-table walk (PTW), identifies the target device, and returns the translation result. To reduce IOMMU pressure, {\large\ding{186}}Trans-FW~\cite{li2023trans} places a centralized forwarding table (FT) at the CPU and accesses it in parallel with the PTW. On an FT hit, the request is sent directly to the GPU holding the requested PTE.

%废案：Only then can the GPU issue the remote access to the target device.

\subsection{Challenge and Keynote}

\textbf{\textit{Challenge:}} CPU-centric translation is fundamentally mismatched to WSGs. Firstly, integrating a CPU onto the same wafer as the GPU array is unachievable because CPUs and GPUs require different process technologies, mask sets, and packaging. Consequently, a translation request forwarded to the CPU-IOMMU must cross the wafer-to-CPU link twice, traversing multiple interconnect hierarchies, incurring substantial latency and congestion.

Secondly, the scalability mismatch further aggravates this cost. A conventional node typically pairs one CPU with eight GPUs, whereas a single wafer can integrate 16--32 GPU dies. As the GPU count grows, remote accesses and translation requests increase, but CPU-side IOTLB capacity and PTW parallelism remain limited. Trans-FW \cite{li2023trans} mitigates some overhead with a forwarding table, but requests still incur the off-wafer transfer and the centralized table can process only one request at a time, becoming a bottleneck under the massive request volume generated by GPU SMs.

\textbf{\textit{Keynote:}} The CPU is the default destination for remote translation because it maintains the complete PTE set, whereas each GPU stores only PTEs for pages resident in its local memory. Yet PTEs for GPU-resident pages are collectively present across the GPU array; the missing piece is not the PTE itself, but an efficient way to locate it.

\begin{keynotebox}
\textbf{Key Insight:} If each GPU can efficiently locate any requested PTE within the array, remote translation can be resolved on the GPUs themselves.
\end{keynotebox}

\section{Design}

\subsection{PTE Presence Consistency (PTE-PC)}

\begin{figure}[t]
    \centering
    \includegraphics[width=1.0\linewidth]{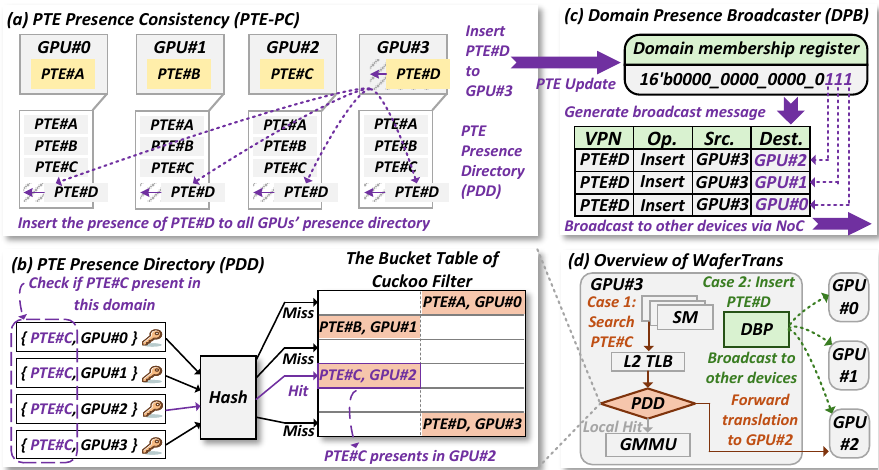}
    \caption{(a) Concept of PTE Presence Consistency (PTE-PC). (b) Architecture of PTE Presence Directory (PPD). (c) Architecture of Domain Presence Broadcaster (DPB). (d) Overview structure of WaferTrans.}
    \label{fig:tech-1}
\end{figure}

We define \textbf{\textit{PTE Presence Consistency (PTE-PC)}} as a lightweight consistency model for translation metadata (Fig.~\ref{fig:tech-1}(a)). A set of GPUs forms a PTE-PC domain if every GPU in the domain has a consistent view of whether a given PTE is present in the domain and, if so, which GPU holds it. Unlike full PTE replication, PTE-PC maintains only compact presence metadata rather than PTE contents. This model is much simpler than conventional memory consistency because GPUs never modify PTE contents directly; PTE updates are generated only by page allocation and migration events. Therefore, PTE-PC needs to track only two operations: insertion, when a PTE enters a domain, and removal, when it leaves.

Fig. \ref{fig:tech-1}(d) illustrates the hardware support. To accommodate the PTE-PC, each GPU is equipped with a \textbf{\textit{PTE Presence Directory (PPD)}}, derived from the forwarding table of Trans-FW~\cite{li2023trans}. While Trans-FW places this directory as a centralized unit in the CPU, we replicate it on every GPU within the same domain. The PPD extends the original cuckoo filter by using the VPN of the PTE concatenated with the \textit{gpu\_id} as the key. On a lookup, the PPD probes $M$ lanes in parallel using the same VPN and all possible \textit{gpu\_id} values. If all lanes miss, the PTE is absent from the domain; otherwise, the matching lane directly identifies the GPU that holds the PTE.

To preserve consistency across replicated PPDs, each GPU is also paired with a \textbf{\textit{Domain Presence Broadcaster (DPB)}} that manages an $N$-bit domain-membership register. This register serves as a bitmask for a cluster of $N$ devices: the $M$ bits for GPUs within this PTE-PC domain are set to $1$, while the remaining $N-M$ bits are cleared to $0$. On a PTE update, the local DPB encapsulates the update and its \textit{gpu\_id} in a broadcast message sent to all domain members; recipients subsequently refresh their local PPDs.

\subsection{Distributed PTE-PC Mapping}

Page migration updates PTEs only at the devices involved in the migration. A straightforward way to expose this information to all GPUs is to build a single global PTE-PC domain and broadcast every PTE update across the entire mesh. This approach is prohibitively expensive. For an $N\times N$ GPU array, each update would trigger chip-wide NoC traffic and require each GPU to maintain $O(N^2)$-scale presence metadata.

\begin{figure}[t]
    \centering
    \includegraphics[width=1.0\linewidth]{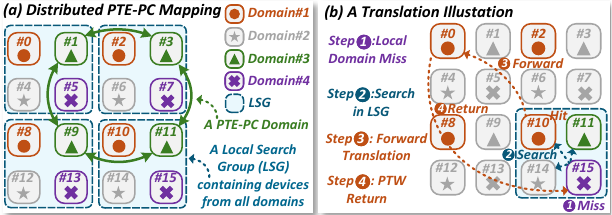}
    \caption{(a) Distributed PTE-PC Mapping. (b) Illustration of remote translation.}
    \label{fig:tech-2}
\end{figure}

To reduce the update and storage cost of PTE-PC, we partition the array into multiple presence domains. Fig.~\ref{fig:tech-2}(a) illustrates a $4\times4$ example, where GPUs of the same color belong to the same domain. For example, the green GPUs $\{G_1, G_3, G_9, G_{11}\}$ form one domain. Whenever a GPU in this domain inserts or removes a PTE, the update is propagated only along the green two-hop ring.

The mapping is deliberately interleaved to support efficient lookup through a second structure, the \textbf{\textit{local search group (LSG)}}. Each LSG contains one GPU from every presence domain; for example, $\{G_{10}, G_{11}, G_{14}, G_{15}\}$ forms an LSG in Fig.~\ref{fig:tech-2}(a). An LSG therefore collectively covers the PTE-presence information for the entire wafer and serves the lookup role formerly performed by the CPU-IOMMU. After a local translation miss, a GPU needs to query only the nearby members of its LSG to locate the requested PTE.

\textbf{\textit{Illustration:}} Fig.~\ref{fig:tech-2}(b) shows a translation example. {\large\ding{182}}Suppose that $G_{15}$ misses locally. {\large\ding{183}}It queries the other members of its LSG, $\{G_{10}, G_{11}, G_{14}\}$. Assume that the target PTE resides at $G_{0}$ in Domain~\#1. {\large\ding{184}}The PPD at $G_{10}$ hits and forwards the translation request to $G_{0}$. {\large\ding{185}}Finally, $G_{0}$ returns the translation result to $G_{15}$.

\section{Evaluation}
\subsection{Evaluation Setup}

\begin{figure}[t]
    \centering
    \includegraphics[width=1.0\linewidth]{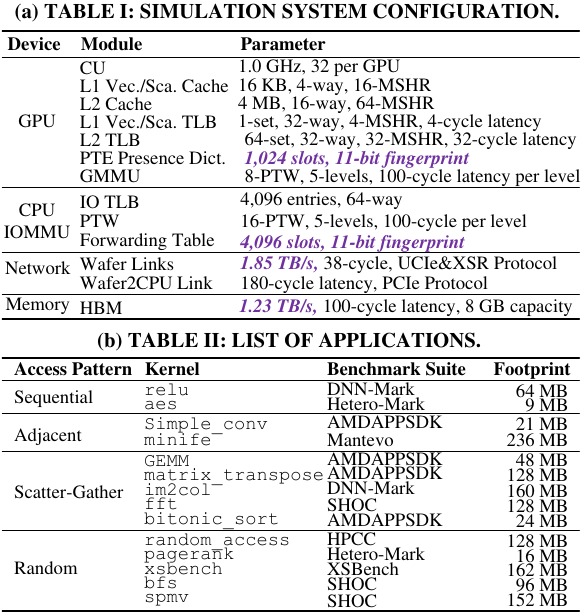}
    \caption{The table of (a) simulation configuration and (b) benchmark selection.}
    \label{fig:tables}
\end{figure}

\textbf{\textit{Simulator:}} 
We use MGPUSim \cite{sun2019mgpusim} to evaluate, which is widely adopted in recent multi-GPU memory-architecture studies. We extend its timing model with WaferTrans. The centralized \texttt{vm.PageTable} is removed from the translation path, and each GPU integrates a \texttt{PPD} and a \texttt{DPB} to locate PTEs and propagate PTE updates within a domain. Finally, we replace the zero-latency \texttt{InterDeviceConn} with a parameterized wafer-scale mesh that models hop latency, bandwidth, routing distance, and queueing contention.

\textbf{\textit{Configuration:}}
We calibrate the single-GPU configuration to an AMD MI100 GPU, as summarized in Fig.\ref{fig:tables}~(a). To keep the simulation tractable, we scale each GPU to one quarter of the original MI100 configuration by reducing the number of CUs and the capacities of the L2 cache and L2 TLB. We configure the wafer links based on the TSMC SoW-X report~\cite{shih2025sow}, which describes a 16-GPU wafer-scale system with 15~TB/s interconnect bandwidth and 10~TB/s memory bandwidth per die. Since the MI100 provides 1.23~TB/s memory bandwidth, we preserve the interconnect-to-memory-bandwidth ratio and set the interconnect bandwidth to $15/10 \times 1.23 = 1.85$~TB/s. CPU-related parameters are applicable only to the baseline.

\textbf{\textit{Benchmarks:}}
Fig.\ref{fig:tables}~(b) lists 14 algorithms we select from, commonly used parallel-workload benchmark suites. These benchmarks cover four representative memory-access patterns: Sequential, Adjacent, Scatter-Gather, and Random.

\subsection{Hardware Cost}
We implement WaferTrans in RTL. The PPD contains 1,024 slots with 11-bit fingerprints, yielding a false-positive rate below 0.39\%. Synthesized in TSMC 28-nm process, the complete design occupies 0.0354~mm$^2$. At the TT corner (0.9~V and 25$^\circ$C) and a frequency of 1~GHz, the complete design consumes 6.011~mW. The PPD fingerprint table logically stores 1,408~Bytes and is mapped to six 128$\times$16-bit 2RW SRAM macros occupying 0.0337~mm$^2$. The remaining DPB and control logic occupy only 0.00167~mm$^2$, only 4.71\% of the total area. Based on the MI100 die shot, 4 MB L2 cache occupies approximately 35.7 mm$^2$; relative to this baseline, WaferTrans introduces only a 0.40\% area overhead.

\begin{figure}[t]
    \centering
    \includegraphics[width=1.0\linewidth]{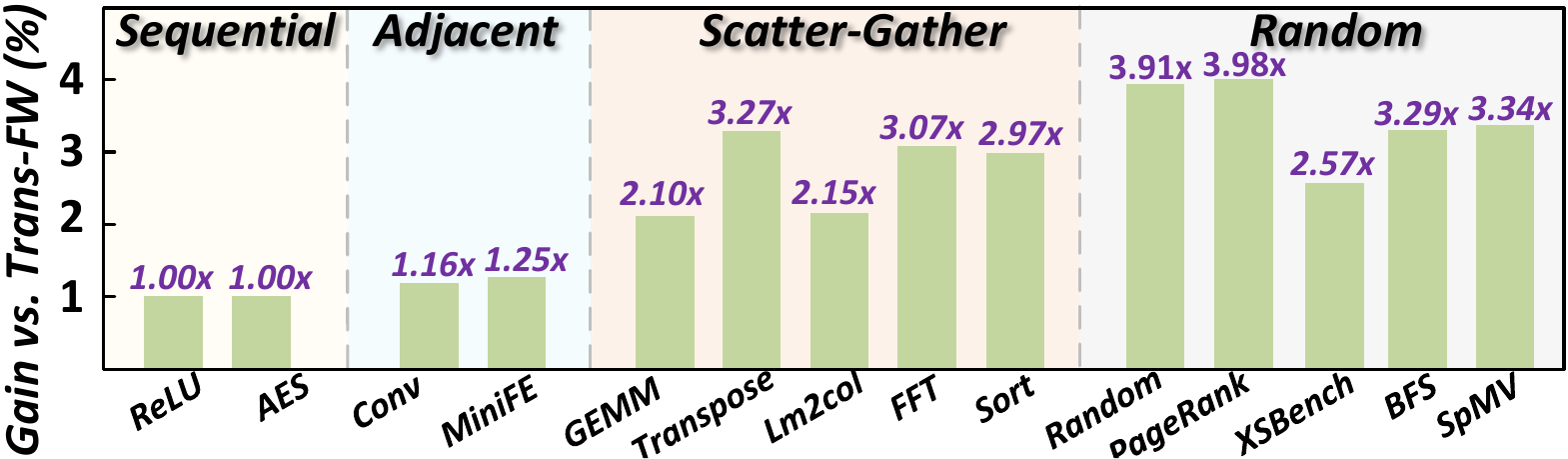}
    \caption{Overall performance comparison of WaferTrans and Trans-FW.}
    \label{fig:exp1-1}
\end{figure}

\begin{figure}[t]
    \centering
    \includegraphics[width=1.0\linewidth]{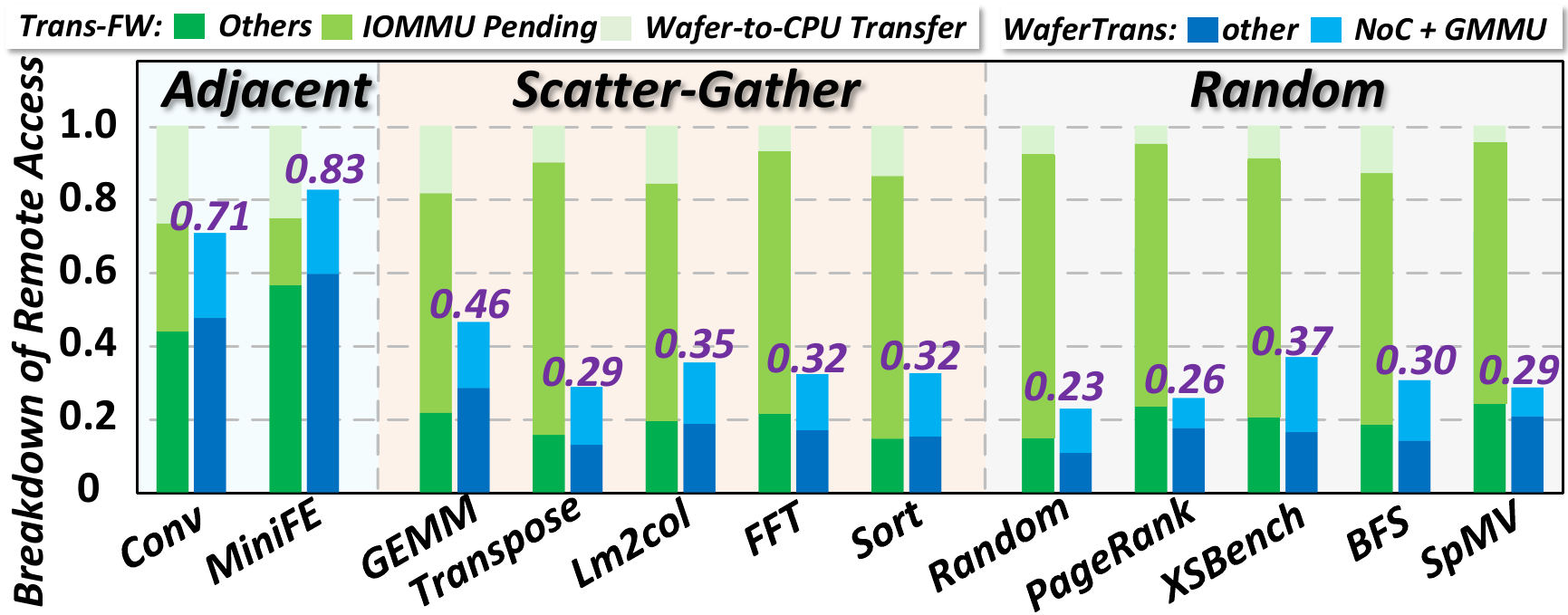}
    \caption{Remote-access latency breakdown of WaferTrans and Trans-FW.}
    \label{fig:exp1-2}
\end{figure}

\begin{figure}[t]
    \centering
    \includegraphics[width=1.0\linewidth]{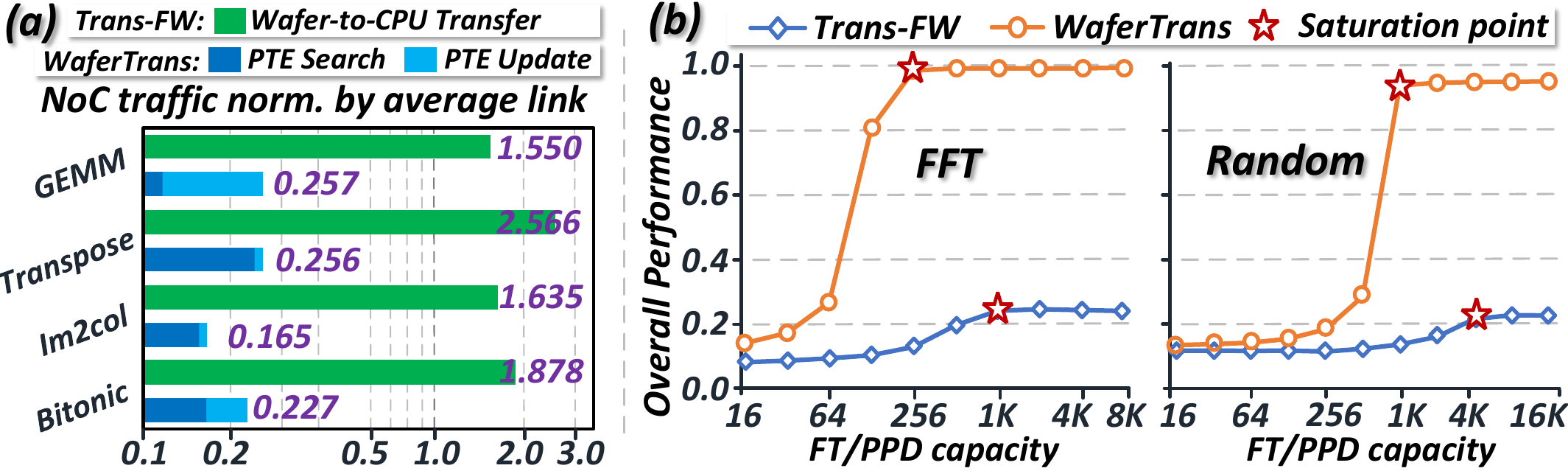}
    \caption{(a) Network pressure and (b) sensitivity to table capacity.}
    \label{fig:exp2-1}
\end{figure}

\subsection{Overall Performance}

Trans-FW~\cite{li2023trans} serves as our baseline. Fig.~\ref{fig:exp1-1} presents the overall performance comparison. WaferTrans outperforms Trans-FW by an average of $2.5\times$. For the Sequential and Adjacent workloads, strong data locality results in few remote accesses, leading to small performance difference. In contrast, Scatter-Gather and Random generate substantial structured and irregular remote-access traffic, respectively. For these workloads, WaferTrans achieves an average speedup of $3.1\times$.

Fig.~\ref{fig:exp1-2} further breaks down the remote-access latency for the workloads that benefit from WaferTrans. In Trans-FW, requests spend most of their time pending in the host IOMMU and traversing the wafer-to-CPU link, which account for average shares of 49\% and 13\% of the total latency, respectively. WaferTrans eliminates the off-wafer transfer overhead and distributes translation requests across the wafer, thereby reducing contention and queueing delays in the translation path.

\subsection{Exploring Network Pressure}

To demonstrate the cost of off-wafer translation and quantify WaferTrans's network overhead, we measure network traffic in Fig.~\ref{fig:exp2-1}(a). The wafer-to-CPU link carries $1.9\times$ the average traffic, making it the primary network bottleneck. In contrast, WaferTrans introduces only two types of traffic: PTE updates and PTE searches. PTE updates are propagated only within the corresponding PTE-PC domain and, because page migrations are infrequent, account for only $6\%$ of the average traffic. PTE searches are broadcast only within local LSGs, whose closely colocated GPUs limit transmission distance, and account for a further $17\%$. Overall, WaferTrans introduces only $23\%$ additional network overhead.

\subsection{Sensitivity to Table Capacity}

The capacities of WaferTrans's PPD and Trans-FW's forwarding table determine how many PTE locations can be stored on chip. Figure~\ref{fig:exp2-1}(b) presents sensitivity results for FFT and RandomAccess. Increasing table capacity improves performance until the table accommodates nearly all PTE locations generated during execution, after which performance saturates. By distributing PTE-location metadata across the wafer, WaferTrans reaches its higher saturated performance with approximately one quarter of the capacity required by Trans-FW. For the most demanding workload, RandomAccess, WaferTrans and Trans-FW saturate at 1{,}024 and 4{,}096 slots, respectively; we use these values as the default configurations in Fig.\ref{fig:tables}~(a).

\section{Conclusion}

This paper presents WaferTrans, an IOMMU-free distributed address translation design for wafer-scale GPUs. By locating PTEs cooperatively within the GPU array, it removes costly CPU-side translation, thus improving performance.

\bibliographystyle{IEEEtran}
\bibliography{refs}

\vfill
\end{document}